\documentclass[pra,aps,twocolumn]{revtex4}
\usepackage{bm,dcolumn,amsmath,graphicx,amsfonts,amssymb,lipsum}
\usepackage{CJK}
\usepackage{longtable}
\usepackage{epsfig}
\begin{document}

	\title{Theoretical study of electronic structure of hafnium (Hf, Z=72)
		and rutherfordium (Rf, Z=104) atoms and their ions: Energy levels and
		hyperfine structure constants}
	\author{Saleh O. Allehabi, V. A. Dzuba, and V. V. Flambaum}

	\affiliation{School of Physics, University of New South Wales, Sydney 2052, Australia}

	
	\date{\today}
	
	\begin{abstract}
	Energy levels, magnetic dipole, and electric quadrupole hyperfine structure of the superheavy element rutherfordium (Rf, $Z$=104) and its first three ions are calculated using a combination of the configuration interaction, coupled-cluster single-doubles, and many-body perturbation theory techniques. The results are to be used in future interpretations of the measurements in terms of nuclear magnetic dipole and electric quadrupole momenta.  To have a guide on the accuracy of the study, we perform similar calculations for hafnium (Hf, $Z$=72) and its ions. Hf is a lighter analog of Rf with a similar electronic structure. Good agreement with the experiment for Hf  and with  available previous calculations of the energy levels of Rf is demonstrated.
	\end{abstract}
	
	\maketitle

\section{Introduction}
	
The study of the hyperfine structure (hfs) of the superheavy elements ($Z > 100$) is a way of obtaining important information about their nuclear structure. The measurements accompanied by atomic calculations lead to extractions of nuclear magnetic dipole and electric quadrupole moments. This serves as a test of nuclear theory leading to more reliable predictions of nuclear properties and helping in the search for the hypothetical \textit{stability island}~\cite{Rev1,Rev2,Rev3,Rev4,Rev5}. The heaviest element so far where such  a study was performed is nobelium (No, $Z=102$)~\cite{No1,No2,No3}. 
The hfs was measured for $^{253}$No isotope in the strong electric dipole transition between ground $^1$S$_0$ and excited $^1$P$^{\rm o}_1$ state.
In addition, isotope shift was measured for the $^{252,253,254}$No isotopes. Similar measurements are now planned for lawrencium (Lr, $Z=103$)~\cite{Lr+}.
Hopefully,  rutherfordium (Rf, Z=104) is next in line.

Most of the synthesized isotopes of Rf have odd neutron numbers~\cite{Rf-IS}, meaning that they have a non-zero nuclear spin and that their energy levels have hyperfine structures. The spectrum of electronic states of Rf was studied theoretically before~\cite{Eliav,Mosyagin,Dzuba2}, revealing several electric dipole transitions suitable for the measurements.

In the present work, we perform calculations of energy levels,
  magnetic dipole, and electric quadrupole hyperfine structure (hfs) of neutral Rf and Hf and their first three ions. The main purpose of the work is to obtain the values of the hfs matrix elements needed for the interpretation of future measurements. The energy levels are obtained as a byproduct; they are also useful for assessing the accuracy of the calculations.
We use a combination of the linearised single-double coupled-cluster method with the configuration interaction technique, the CI+SD method~\cite{Dzuba_SD+CI}.
Calculations for Hf are performed to test the accuracy of the predictions for Rf.
Hf is a lighter analog of Rf with  a similar electronic structure. Good agreement with the experiment for Hf and with previous calculations of the energy levels of Rf is demonstrated. This opens a way for the interpretation of future measurements in terms of the nuclear magnetic dipole and electric quadrupole moments.

\section{Method of calculation}
\subsection{Calculation of energy levels}

For all considered systems, calculations start for the relativistic Hartree-Fock (RHF) procedure for the closed-shell core (Hf~V and Rf~V).
This corresponds to the use of the V$^{N-M}$ approximation~\cite{Dzuba1}. Here $N=Z$ is the total number of electrons in a neutral atom, and $M$ is the number of valence electrons ($M=4$ for Hf~I and Rf~I).
The RHF Hamiltonian has the form
\begin{equation} \label{e:RHF}
	\hat H^{\rm RHF}= c\alpha\hat p+(\beta -1)mc^2+V_{\rm nuc}(r)+V_{\rm core}(r),
\end{equation}
where $c$ is the speed of light, $\alpha$ and $\beta$ are the Dirac matrixes, $\hat p$ is the electron momentum, $m$ is the electron mass, $V_{\rm nuc}$ is the nuclear potential obtained by integrating Fermi distribution of nuclear charge density, $V_{\rm core}(r)$ is the self-consistent RHF potential created by the electrons of the closed-shell core. 

After the self-consistent procedure for the core is completed, the full set of single-electron states is generated using the B-spline technique~\cite{Johnson_Bspline,Johnson_Bspline2}. The basis states are linear combinations of B-splines, which are the eigenstates of the RHF Hamiltonian (\ref{e:RHF}). We use 40 B-splines of the order  9 in a box that has a radius R$_{\rm max}$ = 40$a_B$ with the orbital angular momentum 0~$\leq$~\textit{l}~$\leq$~6. These basis states are used for solving the linearised single-double couple-cluster (SD) equations and for generating many-electron states for the configuration interaction (CI) calculations. By solving the SD equations first for the core and then for the valence states, we obtain correlation operators $\Sigma_1$ and $\Sigma_2$~\cite{Dzuba_SD+CI}. $\Sigma_1$ is a one-electron operator which is responsible for correlation interaction between a particular valence electron and the core. $\Sigma_2$ is a two-electron operator that can be understood as screening of Coulomb interaction between a pair of valence electrons by core electrons. These $\Sigma$ operators can be used in the subsequent  CI calculations for atoms with several valence electrons to account for the core-valence and core-core correlations.
Solving the SD equations for valence states also gives energies of the single-electron states for the system with one external electron above closed shells. Note that there is a small difference in the SD equations intended for obtaining these energies compared to those intended for further use in the CI calculations. In the latter case, one particular term should be removed from the SD equations since its contribution is included via the CI calculations (see Ref.~\cite{Dzuba_SD+CI} for details). However, the contribution of this term is small, and the difference in the SD equations can be neglected.

The CI equations
\begin{equation} \label{e:CI}
\langle a | \hat H^{\rm CI} |b \rangle - E \delta_{ab} = 0
\end{equation}
have the CI Hamiltonian, which includes $\Sigma_1$ and $\Sigma_2$,
\begin{eqnarray} \label{e:HCI}
	\hat H^{\rm CI}&=&\sum_{i=1}^{M} \left(\hat H^{\rm RHF}+\Sigma_1\right)_i \\
	                  &+&\sum_{i<j}^{M} \left(\dfrac{e^2}{|r_i-r_j|}+ \Sigma_{2ij}\right).\nonumber 
\end{eqnarray}
$a$ and $b$ in (\ref{e:CI}) are many-electron single determinant basis states which are constructed by exciting one or two electrons from one or more reference configuration(s) and then building from these configurations the states of definite values of the total momentum $J$. 
$M$ in (\ref{e:HCI}) is the number of valence electrons. In our cases $M=1,2,3,4$. The case of one external electron is a special one. 
It has no terms of the second line in the CI Hamiltonian (\ref{e:HCI}). Taking into account that single-electron basis states are eigenstates of the RHF
Hamiltonian (\ref{e:RHF}), the CI eigenvalue problem is reduced to diagonalization of the $\Sigma_1$ matrix,
\begin{equation} \label{e:CI1}
\langle i | \Sigma_1 |j \rangle - E \delta_{ij} = 0.
\end{equation}
Here $i$ and $j$ are single-electron basis states.
Note that in spite of significant simplifications of the CI equations for $M=1$, there is no need for the modification of the computer code.
For $M=1$ Eqs. (\ref{e:CI}) and (\ref{e:CI1}) are equivalent.

There is an alternative way to perform the calculations for systems with one external electron.
One can find the energies and wave functions of the valence states by solving the RHF-like equations for an external electron in which the correlation potential $\Sigma_1$ is included,
\begin{equation}\label{e:BO}
\left( \hat H^{\rm RHF} + \Sigma_1 -\epsilon_v \right) \psi_v =0.
\end{equation}
Here index $v$ numerate states of an external electron, wave functions $\psi_v$ are usually called Brueckner orbitals (BO)~\cite{Dzuba_BO}, the energies $\epsilon_v$ and wave functions $\psi_v$ include correlation corrections. BO can be used to calculate matrix elements, in particular, for the hfs (see below). 
Comparing two ways of the calculations is an important test of the accuracy. It is especially valuable when there is  a lack of experimental data, which is the case of the present work.

\begin{figure}[tb]
	\epsfig{figure=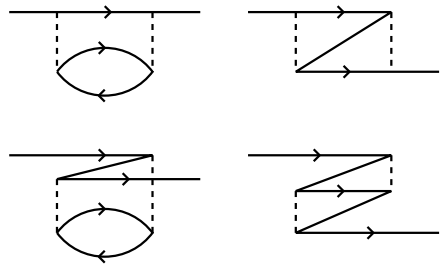,scale=0.5}
	\caption{Four diagrams for the second-order correlation operator $\Sigma_1$.}
	\label{f:sigma}
\end{figure}

The meaning of the $\Sigma_1$ operator is the same in the CI and BO equations (\ref{e:CI1}), (\ref{e:BO}). However, the $\Sigma_1$ operator, which comes from the SD calculations as a set of matrix elements between single-electron states, cannot be directly used in (\ref{e:BO}) since here we need the operator in the coordinate representation. Therefore, we calculate $\Sigma_1$ for the BO in the second-order of the many-body perturbation theory (see Fig.~\ref{f:sigma}). A particular class of the higher-order correlations is included by solving Eq. (\ref{e:BO}) iteratively. It includes contributions $\sim \Sigma_1^2$, $\Sigma_1^3$, etc. In the end, the two ways of calculations are sufficiently different to be a good test of accuracy.

\subsection{The CIPT~Method}

It is very well known that the size of the CI matrix grows exponentially with the number of valence electrons. In  the present work, we have up to four valence electrons (in neutral Hf and Rf), leading to the huge size of the CI matrix, the number of lines in (\ref{e:CI}) $\sim 10^6$. Dealing with a matrix of this size requires significant computer power. However, it can be reduced by orders of magnitude for the cost of some sacrifices in the accuracy of the result. 
To do this, we use the CIPT method \cite{CIPT} (configuration interaction with perturbation theory). The idea is to neglect off-diagonal matrix elements between high-energy states in the CI matrix (since in the perturbation theory approach such matrix elements appear in higher orders).
Then the CI matrix equation (\ref{e:CI}) can be written in a block form
\begin{equation} \label{e:blocks}
\left( \begin{array}{cc} A & B \\ C & D \end{array} \right) \left(\begin{array}{c} X \\ Y \end{array} \right) = E_a \left(\begin{array}{c} X \\ Y \end{array} \right).
\end{equation}
Here block $A$ corresponds to low-energy states, block $D$ corresponds to high-energy states and~blocks $B$ and $C$ correspond to cross terms. 
Note that since the total CI matrix is symmetric, we have $C = B'$, i.e.,~$c_{ij} = b_{ji}$. 
Vectors $X$ and $Y$ contain the coefficients of expansion of the valence wave function over single-determinant many-electron basis functions,
\begin{eqnarray}
&&\Psi (r_1, \dots ,r_M)=   \label{e:Psi}  \\
&&\sum_{i=1}^{N_1} x_i \Phi_i (r_1, \dots ,r_M)+ \sum_{j=1}^{N_2} y_j \Phi_j(r_1, \dots ,r_M). \nonumber
\end{eqnarray}
Here $M$ is the number of valence electrons, $N_1$ is the number of low-energy basis states, $N_2$ is the number of high-energy basis states.

We neglect off-diagonal matrix elements in block $D$.
Finding $Y$ from the second equation of (\ref{e:blocks}) leads to
\begin{equation}\label{e:Y}
Y=(E_vI-D)^{-1}CX.
\end{equation}
Substituting $Y$ to the first equation of (\ref{e:blocks}) leads to
\begin{equation}\label{e:CIPT}
\left[A + B(E_aI-D)^{-1}C\right] X = E_v X,
\end{equation}
where $I$ is  the unit matrix. Neglecting the off-diagonal matrix elements in $D$ leads to a very simple structure of the $(E_aI-D)^{-1}$ matrix, $(E_aI-D)^{-1}_{ik} = \delta_{ik}/(E_a - E_k)$, where $E_k = \langle k|H^{\rm CI} |k \rangle$ (see~\cite{CIPT} for more~details). The relative sizes of blocks $A$ and $D$ can be varied in the calculations in search for a reasonable compromise between the accuracy of the results and  the computer power needed to obtain them.
In our current calculations, the number of lines in (\ref{e:CIPT}) is $\sim 10^3$.

\subsection{Calculation of hyperfine structure}

To calculate hfs,
 we use the time-dependent Hartree-Fock (TDHF) method, which is equivalent to the well-known random-phase approximation (RPA).
The RPA equations are the following:
	
\begin{equation}\label{e:RPA}
	\left(\hat H^{\rm RHF}-\epsilon_c\right)\delta\psi_c=-\left(\hat f+\delta V^{f}_{\rm core}\right)\psi_c
\end{equation}
where $\hat f$ is an operator of  an external field (external electric field, nuclear magnetic dipole or electric quadrupole fields).  
Index $c$ in (\ref{e:RPA}) numerates states in the core, $\psi_c$ is a single-electron wave function of the state $c$ in the core, $\delta\psi_c$ is the correction to this wave function caused by an external field, and $\delta V^{f}_{\rm core}$ is the correction to the self-consistent RHF potential caused by changing of all core states. The nucleus is assumed to be a sphere with a uniform distribution of the tnuclear electric quadrupole moment and nuclear magnetic dipole moment. Eqs. (\ref{e:RPA}) are solved self-consistently for all states in the core. As a result, an effective operator of the interaction of valence electrons with an external field is constructed as $\hat f + \delta V^{f}_{\rm core}$. Energy shift of a many-electron state $a$, which is a solution of the CI equations (\ref{e:CI}), is given by
\begin{equation} \label{e:de}
\delta \epsilon_a = \langle a | \sum_{i=1}^M \left(\hat f+\delta V^f_{\rm core} \right)_i | a\rangle.
\end{equation}
When the wave function for the valence electrons comes as a solution of Eq.~(\ref{e:CIPT}), Eq.~(\ref{e:de}) is reduced to
\begin{equation}\label{e:mex}
\delta \epsilon_a = \sum_{ij} x_i x_j \langle \Phi_i|\hat H^{\rm hfs}|\Phi_j \rangle,
\end{equation}
where $\hat H^{\rm hfs} =  \sum_{i=1}^M (\hat f+\delta V^f_{\rm core})_i$.
For better accuracy of the results, the full expansion (\ref{e:Psi}) might be used. Then it is convenient to introduce  a new vector $Z$, which contains both $X$ and $Y$, $Z \equiv \{X,Y\}$. Note that the solution of (\ref{e:CIPT}) is normalized by the condition $\sum_i x_i^2=1$. The normalization condition for the total wave function (\ref{e:Psi}) is different,  $\sum_i x_i^2+\sum_j y_j^2 \equiv \sum_i z_i^2=1$. Therefore, when $X$ is found from (\ref{e:CIPT}), and $Y$ is found from (\ref{e:Y}), both vectors should be renormalized. Then the hfs matrix element is given by the expression, which is similar to (\ref{e:mex}) but has much more terms 
\begin{equation}\label{e:mez}
\delta \epsilon_a = \sum_{ij} z_i z_j \langle \Phi_i|\hat H^{\rm hfs}|\Phi_j \rangle.
\end{equation}

In the case of one external electron, the calculations can also be done using BO,
\begin{equation} \label{e:dv}
\delta \epsilon_v = \langle v | \hat f+\delta V^f_{\rm core}  | v\rangle.
\end{equation}
Here $v$ stands for a solution of the Eq. (\ref{e:BO}). 
Energy shifts (\ref{e:de}), (\ref{e:dv}) are used to calculate hfs constants $A$ and $B$ using textbook formulas
\begin{equation}
A_a = \frac{g_I \delta \epsilon_a^{(A)}}{\sqrt{J_a(J_a+1)(2J_a+1)}},
\label{e:Ahfs}
\end{equation}
and
\begin{equation}
B_a = -2Q \delta \epsilon_a^{(B)}\sqrt{\frac{J_a(2J_a-1)}{(2J_a+3)(2J_a+1)(J_a+1)}}. 
\label{e:Bhfs}
\end{equation}
Here $\delta \epsilon_a^{(A)}$ is the energy shift (\ref{e:de}) or (\ref{e:dv}) caused by the interaction of atomic electrons with the nuclear magnetic moment $\mu$, $g_I=\mu/I$, $I$ is nuclear spin; $\delta \epsilon_a^{(B)}$ is the energy shift (\ref{e:de}) or (\ref{e:dv}) caused by the interaction of atomic electrons with the nuclear electric quadrupole moment $Q$ ($Q$ in (\ref{e:Bhfs}) is measured in barn). 

\subsection{Further corrections to the hyperfine structure.}

\label{s:Ba+}
Using Eq.~(\ref{e:mex}) is the fastest way of calculating hfs for many-electron atoms. Sometimes it gives pretty accurate results, within $\sim$~10\% of the experimental values. This is usually the case when the hfs comes mostly from contributions of the $s$ and $p$ states. In our case, the contribution of the $s$ states is suppressed because in the leading configurations ($6s^25d^2$ and $6s^25d6p$), the $6s$ electrons are from the closed subshell and do not contribute to the hfs. This means that further corrections to the hfs matrix elements should  be considered. The Eq.~(\ref{e:mex}) can still be used to identify states with large hfs. The accuracy of the calculations is likely to be higher for such states. This is because the small value of hfs often comes as a result of strong cancellations between different contributions leading to poor accuracy of the results.

\begin{figure}[tb]
	\epsfig{figure=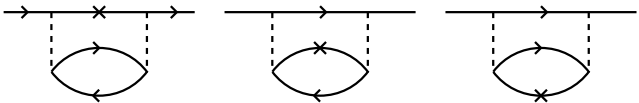,scale=0.5}
	\caption{Sample SR diagrams corresponding to the first diagram on Fig.~\ref{f:sigma}. Cross stands for the hfs operator. It goes to all internal lines of all four diagrams for the $\Sigma_1$ operator.}
	\label{f:SR}
\end{figure}

\begin{figure}[tb]
	\epsfig{figure=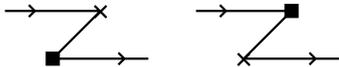,scale=0.5}
	\caption{Self-energy diagrams. Cross stands for the hfs operator, black box stands for the correlation operator $\Sigma_1$ (see Fig.~\ref{f:sigma}).}
	\label{f:SE}
\end{figure}

\begin{figure}[tb]
	\epsfig{figure=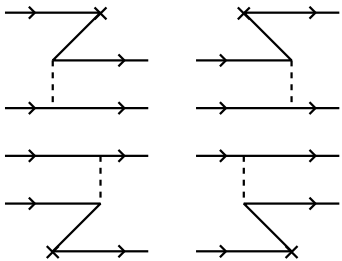,scale=0.5}
	\caption{Two - particle correction to the many-electron matrix element of the hfs interaction. Cross stands for the hfs operator, dashed line is the Coulomb interaction.}
	\label{f:TP}
\end{figure}

There are at least three classes of the higher-order corrections to the hfs matrix elements: (a) Contribution of the higher states (HS). This is the difference between  (\ref{e:mez}) and (\ref{e:mex}). (b) Corrections to the single-electron matrix elements caused by the correlation operator $\Sigma_1$. 
This includes {\em the structure radiation} (SR) when the hfs operator is inside of the $\Sigma_1$ operator (see Fig.~\ref{f:SR}), and the self-energy correction, 
when the hfs operator is outside of the $\Sigma_1$ operator (see Fig.~\ref{f:SE}) (see also \cite{JETP1,JETP2}). (c) Two-particle correction \cite{JETP1,JETP2}, 
which is a correction to the Coulomb interaction between valence electrons caused by the hfs interaction (see Fig.~\ref{f:TP}).

To study the corrections to the single-electron matrix elements, it is convenient to have a system with one external electron above closed shells where experimental data are available for a range of valence states. The $^{135}$Ba$^+$ ion is a good example of such a system. Table~\ref{t:Ba+} presents  a comparison with the measured magnetic dipole hfs constants of $^{135}$Ba$^+$, calculated in different approximations. The RHF (relativistic Hartree-Fock) column corresponds to using Eq.~(\ref{e:dv}) in which valence state $|v\rangle$ is s Hartree-Fock orbital and core polarization correction $\delta V^f_{\rm core}$ is absent, the RPA column, in the RPA column the CP correction is added, BORPA corresponds to using BO in (\ref{e:dv}), the BORPA rescaled is the same but BO calculated with rescaled correlation operator $\lambda\Sigma_1$; the rescaling parameter $\lambda$ is chosen to fit experimental energies. The SR column is structure radiation (Fig.~\ref{f:SR}), the {\em total} column is the sum of the previous two columns. The last column presents the experimental hfs constants from Ref.~\cite{Ba+expt1,Ba+expt2}. The table shows that all considered corrections are important, and including them all leads to
 accurate results in most cases. Therefore, all these corrections should be included in the calculations for many-electron atoms via correcting single-electron matrix elements. Inclusion of the CP and SR corrections is straightforward, but  the inclusion of correlations like in BO needs some clarification. The CI Hamiltonian (\ref{e:HCI}) does include the correlation operator $\Sigma_1$ leading to  the  mixing of the states above the core and forming orbitals similar to BO. No core states are involved in this mixing.  On the other hand, the BO found by solving the Eq.~(\ref{e:BO}) can be written in  the first order of $\Sigma_1$ as
 \begin{equation}\label{e:BO1}
 \psi_v^{\rm BO} = \sum_i |i\rangle\frac{\langle i|\Sigma_1|v\rangle}{E_v-E_i}.
 \end{equation}
 Here summation goes over  the complete set of the single-electron Hartree-Fock states, including states in the core. The self-energy (SE) terms (Fig.~\ref{f:SE}) are needed to account for this missed summation over  the core states in the CI calculations.
\begin{table}
\caption{Magnetic dipole hfs constants of $^{135}$Ba$^+$ (MHz) calculated in different approximations.}
\label{t:Ba+}
\begin{ruledtabular}
\begin{tabular}{l rrrrrrr}
\multicolumn{1}{c}{State}&
\multicolumn{1}{c}{RHF}&
\multicolumn{1}{c}{RPA}&
\multicolumn{1}{c}{BORPA}&
\multicolumn{1}{c}{BORPA}&
\multicolumn{1}{c}{SR}&
\multicolumn{1}{c}{Total}&
\multicolumn{1}{c}{Expt.}\\
&&&&\multicolumn{1}{c}{rescaled}&&&
\multicolumn{1}{c}{\cite{Ba+expt1,Ba+expt2}}\\
\hline
$6s_{1/2}$ & 2603 & 3090 & 3815 & 3654 & -63 & 3591 & 3593  \\
$6p_{1/2}$ &  440 &  530 &  691 &  659 &   5 &  664 &  665  \\
$6p_{3/2}$ &   64 &  105 &  134 &  129 & -16 &  113 &  113  \\
$5d_{3/2}$ &  115 &  133 &  165 &  161 &  27 &  188 &  170  \\
$5d_{5/2}$ &   46 &  -50 &  -47 &  -48 &  35 &  -13 & -10.7 \\ 
\end{tabular}
\end{ruledtabular}
\end{table}
Table~\ref{t:Ba+} shows that the considered approximation gives very accurate results for the hfs of $s$ and $p$ state, while the results for $d$ states are less accurate. Furthermore, the relative difference between theory and experiment for $d_{3/2}$ states is about two times smaller than for the $d_{5/2}$ states.
This means that while considering hfs of many-electron atoms in which the values of the hfs constants come mostly from the contribution of the $d$ states (like Hf and Rf), it is preferable to consider states in which the contribution of the $d_{3/2}$ states dominates over the contribution of the $d_{5/2}$ states.
The accuracy of the calculations is likely to be higher for these states. To identify such states, we need to do the analysis of the partial contributions to the hfs  of the many-electron atoms. We will further discuss the matter in section~\ref{s:PC}.

\section{Results}

\subsection{Energy levels of Hf, Rf and their ions.}
\begin{table}
	\caption{\label{t:HfEL}
		Excitation energies ($\textit{E}$, cm$^{-1}$)
		for some low states of Hf~I, Hf~II, Hf~III, and Hf~IV.}
	\begin{ruledtabular}
		\begin{tabular}{cllccc}
			&&&&
			\multicolumn{1}{c}{ Present }&
			\multicolumn{1}{c}{ Expt.}\\
			
			\cline{5-5}
			
			\multicolumn{1}{c}{No.}& 
			\multicolumn{1}{c}{Conf.}&
			\multicolumn{1}{c}{Term}&
			
			\multicolumn{1}{c}{\textit{J}}&
			\multicolumn{1}{c}{(CI+SD)}&\\
			
			\hline
			
			\multicolumn{5}{c}{\textbf{Hf~I}}&
			\multicolumn{1}{c}{\cite{NIST}}\\
			
			1 & $5d^{2}6s^2$& $^3${F}& 2 &0 &0 \\
			2 & $5d^{2}6s^2$& $^3${F}& 3 &2114&2356.68\\
			3& $5d^{2}6s^2$& $^3${F}& 4 &4148&4567.64\\
			4 & $5d^{2}6s^2$& $^1${D}& 2 & 4799&5638.62\\
			5& $5d^{2}6s^2$& $^3${P}& 1 &5063&6572.54\\
			6& $5d^{2}6s^2$& $^3${P}& 2 &9026&8983.75\\
			7& $5d6s^26p$& $^1${D}$\rm ^o$& 2 &10634&10508.88\\
			8& $5d^{2}6s^2$& $^1${G}& 4 &10402&10532.55\\
			9& $5d6s^26p$& $^3${D}$\rm ^o$& 1 &14042&14017.81\\
			10& $5d^{3}6s$& $^5${F}& 1 &12469&14092.26\\
			11& $5d6s^26p$& $^3${F}$\rm ^o$& 2 &14092&14435.12\\
			12& $5d6s^26p$& $^3${F}$\rm ^o$& 3 &14545&14541.66\\
			13& $5d^{3}6s$& $^5${F}& 2 &12625&14740.67\\
			14& $5d^{3}6s$& $^5${F}& 3 &13181&15673.32\\
			15& $5d6s^26p$& $^3${D}$\rm ^o$& 2 &15706&16163.35\\	
			16& $5d^{3}6s$& $^5${F}& 4 &14050&16766.60\\
			17& $5d^26s6p$& $^5${G}$\rm ^o$& 2 &18234&18011.04\\	
			18& $5d6s^26p$& $^3${P}$\rm ^o$& 1 &17969&18143.39\\
			19& $5d6s^26p$& $^3${F}$\rm ^o$& 4 &16485&18224.97\\
			20& $5d6s^26p$& $^3${D}$\rm ^o$& 3 &17824&18381.50\\
			21& $5d^26s6p$& $^5${G}$\rm ^o$& 3 &19148&19292.68\\
			22& $5d6s^26p$& $^3${P}$\rm ^o$& 2 &19490&19791.29\\
			23& $5d^{3}6s$& $^5${P}& 1 &18363&20784.87\\
			24& $5d^{3}6s$& $^5${P}& 2 &19085&20908.43\\
			\hline
			
			\multicolumn{5}{c}{\textbf{Hf~II}}&
			\multicolumn{1}{c}{\cite{NIST}}\\
			
			1 & $5d6s^2$& $^2${D}& {3/2} &0 &0\\
			2 & $5d6s^2$& $^2${D}& {5/2} &3054&3050.88\\
			3& $5d^{2}6s$& $^4${F}& {3/2} &3578&3644.65\\
			
			4& $5d^{2}6s$& $^4${F}& {5/2} &4312&4904.85\\
			5 & $5d^{2}6s$& $^4${F}& {7/2} &5330&6344.34\\
			6 & $5d^{2}6s$& $^4${F}& {9/2} &8039&8361.76\\
			7& $5d^{2}6s$& $^4${P}& {1/2} &11675&11951.70\\
			8& $5d^{2}6s$& $^2${F}& {5/2} &11783&12070.46\\
			9& $5d^{2}6s$& $^4${P}& {3/2} &11781&12920.94\\
			10& $5d^{2}6s$& $^4${P}& {5/2} &12581&13485.56\\
			
			11& $5d^{2}6s$& $^4${D}& {3/2} &13836&14359.42\\
			12& $5d^{2}6s$& $^2${F}& {7/2} &14410&15084.26\\

			13& $5d6s6p$& $^4${F}$\rm ^o$& {3/2} &28580&28068.79\\
			14& $5d6s6p$& $^4${D}$\rm ^o$& {1/2} &29249&29160.04\\
			15& $5d6s6p$& $^4${F}$\rm ^o$& {5/2} &29759&29405.12\\
			16 & $5d6s6p$& $^4${D}$\rm ^o$& {3/2} &31903&31784.16\\
			
			\hline
			
			\multicolumn{5}{c}{\textbf{Hf~III}}&
			\multicolumn{1}{c}{\cite{Klinkenberg}}\\
			
			1 & $5d^{2}$& $^3${F}& 2 &0 &0 \\
			2 & $5d6s$& $^3${D}& 2 &2572&3039.7\\
			3& $5d^{2}$& $^3${F}& 3 &1944&3288.7\\
			
			4& $5d^{2}$& $^1${D}& 2 &5212&5716\\
			5 & $5d^{2}$& $^3${F}& 4 & 5598&6095.1\\
			6 & $5d6s$& $^3${D}& 3 &6443&6881.6 \\
			7& $5d^{2}$& $^3${P}& 2 &11909&12493.2\\
			
			\hline
			
			\multicolumn{5}{c}{\textbf{Hf~IV}}&
			\multicolumn{1}{c}{\cite{Klinkenberg}}\\
			
			1 & $4f^{14}5d$& $^2$D&{3/2} &0&0\\
			2 & $4f^{14}5d$& $^2${D}&{5/2} &4721 &4692\\
			3 &$4f^{14}6s$&$^2${S}&{1/2} &17530&18380\\
			4 & $4f^{14}6p$& $^2${P}$\rm ^o$&{1/2} &66611&67039\\
			5 & $4f^{14}6p$& $^2${P}$\rm ^o$&{3/2} &76232&76614\\
			6 & $4f^{14}7s$&$^2${S}&{1/2} &140329&140226\\
			
		\end{tabular}
	\end{ruledtabular}
\end{table}

\begin{table}[!]
	
	\caption{\label{t:RfEL}
		Excitation energies ($\textit{E}$, cm$^{-1}$)  for some low states of Rf~I, Rf~II , Rf~III, and Rf~IV.}
	\begin{ruledtabular}
		\begin{tabular}{cllccccc}
			&&&&
			\multicolumn{1}{c}{ Present}&
			\multicolumn{3}{c}{ Other Cal.}\\
			\cline{5-5}
			\cline{6-8}
			\multicolumn{1}{c}{No.}& 
			\multicolumn{1}{c}{Conf.}&
			\multicolumn{1}{c}{Term}&
			\multicolumn{1}{c}{\textit{J}}&
			
			\multicolumn{1}{c}{ (CI+SD)}\\
			
			\hline
			
			\multicolumn{5}{c}{\textbf{Rf~I}}&
			\multicolumn{1}{c}{\cite{Eliav}}&
			\multicolumn{1}{c}{\cite{Mosyagin}}&
			\multicolumn{1}{c}{\cite{Dzuba2}}\\
			
			1  &$7s^{2}6d^{2}$ & $^3${F}     &2 & 0&0&0&0\\
			2  &$7s^{2}7p6d$  & $^3${F}$^{o}$&2 & 2737  & 2210 & 3923 &  2547  \\
			3  &$7s^{2}6d^{2}$ & $^3${F}     &3 & 3836  & 4855 & 4869 &  4904  \\
			4  &$7s^{2}6d^{2}$ & $^3${P}     &2 & 6565  & 7542 & 8704 &  7398  \\
			5  &$7s^{2}6d^{2}$ & $^3${P}     &1 & 7294  & 8776 &10051 &  8348  \\
			6  &$7s^{2}6d^{2}$ & $^3${F}     &4 & 7836  & 7542 & 8597 &  8625  \\
			7  &$7s^{2}7p6d$  & $^3${D}$^{o}$&1 & 8028  & 8373 & 9201 &  8288  \\
			8  &$7s^{2}7p6d$  & $^3${D}$^{o}$&2 &11235  &10905 &12889 & 11273  \\
			9  &$7s^{2}7p6d$  & $^3${F}$^{o}$&3 &11328  &11905 &12953 & 11390  \\
			10 &$7s^{2}7p6d$  & $^1${D}$^{o}$&2 &13811  &$-$   &$-$   & 14403  \\
			11 &$7s^{2}6d^{2}$ & $^1${D}     &2 &13841  &$-$   &$-$   & 13630  \\
			12 &$7s^{2}6d^{2}$ & $^1${G}     &4 &14040  &$-$   &$-$   & 14476  \\
			13 &$7s^{2}7p6d$  & $^3${P}$^{o}$&1 &16017  &$-$   &$-$   & 16551  \\
			14 &$7s^{2}7p6d$  & $^3${D}$^{o}$&3 &17367  &$-$   &$-$   & 18029  \\
			15 &$7s^{2}7p6d$  & $^3${F}$^{o}$&4 &19979  &$-$   &$-$   & 20477  \\
			16 &$7s6d^{2}7p$  & $^5${G}$^{o}$&2 &20371  &$-$   &$-$   & 20347  \\
			17 &$7s6d^{3}$    & $^5${F}     &1 &20626  &$-$   &$-$   & 21552  \\
			18 &$7s^{2}7p6d$  & $^3${P}$^{o}$&2 &21031  &$-$   &$-$   & 21480  \\
			19 &$7s6d^{3}$    & $^5${F}     &2 &21512  &$-$   &$-$   & 23079  \\
			20 &$7s6d^{2}7p$  & $^5${G}$^{o}$&3 &22941  &$-$   &$-$   & 23325  \\
			21 &$7s6d^{3}$    & $^5${F}     &3 &23002  &$-$   &$-$   & 25432  \\
			22 &$7s^{2}7p6d$  & $^1${F}$^{o}$&3 &23965  &$-$   &$-$   & 24634  \\
			23 &$7s6d^{3}$    & $^5${F}     &4 &25231  &$-$   &$-$   & $-$    \\
			24 &$7s6d^{2}7p$  & $^5${F}$^{o}$&1 &25821  &$-$   &$-$   & $-$    \\
			\hline
			
			\multicolumn{5}{c}{\textbf{Rf~II}}&
			\multicolumn{1}{c}{\cite{Eliav}}&
			\multicolumn{1}{c}{\cite{Rf+}}&\\
			
			1  & $7s^{2}6d$& $^2${D}  &{3/2} &0        &   0   &0&$-$\\
			2  & $7s^{2}6d$& $^2${D}  &{5/2} & 7026    & 7444 & 5680&$-$\\
			3  & $7s6d^{2}$& $^4${F}  &{3/2} &15030    &   $-$   &15678&$-$\\
			4  & $7s6d^{2}$& $^4${F}  &{5/2} &16817    &  $-$    &17392&$-$\\
			5  & $7s^{2}7p$& $^2$P$^o$&{1/2} &19050    &19390 &16657&$-$\\
			6  & $7s6d^{2}$& $^4${P}  &{1/2} &23701    &   $-$   &24615&$-$\\
			7  & $7s6d^{2}$& $^4${D}  &{5/2} &25392    &      $-$&26565&$-$\\
			8  & $7s6d^{2}$& $^4${P}  &{3/2} &25561    &      $-$&26648&$-$\\
			9  & $7s6d^{2}$& $^4${D}  &{3/2} &28940    &      $-$&29983&$-$\\
			10 & $7s7p6d$ & $^4$F$^o$ &{3/2} &30264    &      $-$&27846&$-$\\
			11 & $7s6d^{2}$& $^2${P}  &{1/2} &31238     &     $-$&32550&$-$\\
			12 & $7s7p6d$ & $^4$F$^o$ &{5/2} &33320    &      $-$&31031&$-$\\
			13 & $7s^{2}7p$& $^2$P$^o$ &{3/2} &33621    &35513 &31241&$-$\\
			14 & $7s7p6d$ & $^4$D$^o$ &{1/2} &37378    &      $-$&36156&$-$\\
			15 & $7s7p6d$ & $^2$P$^o$ &{3/2} &40015    &      $-$&38814&$-$\\
			16 & $7s7p6d$ & $^4$D$^o$ &{5/2} &40640    &    $-$  &42410& $-$\\
			\hline
			
			\multicolumn{8}{c}{\textbf{Rf~III}}\\
			
			1 & $7s^{2}$& $^1${S} &{0} &0 &$-$&$-$&$-$\\
			2 & $7s6d$& $^3${D} &{1} &8526&$-$&$-$&$-$\\
			3 & $7s6d$& $^3${D} &{2} &9945&$-$&$-$&$-$ \\
			4 & $7s6d$& $^3${D} &{3} &16878&$-$&$-$&$-$\\
			5 & $7s6d$& $^1${D} &{2} &19165&$-$&$-$&$-$\\
			6& $6d^2$& $^3${F} &{2} &24371&$-$&$-$&$-$\\
			7& $6d^2$& $^3${F} &{3} &28326&$-$&$-$&$-$\\
			
			\hline
			
			\multicolumn{8}{c}{\textbf{Rf~IV}}\\
			
			1 &$5f^{14}7s$&$^2${S} &{1/2} &0 &$-$&$-$&$-$\\
			2 & $5f^{14}6d$& $^2${D}&{3/2} &3892&$-$&$-$&$-$\\
			3 & $5f^{14}6d$&$^2${D} &{5/2} &13559&$-$&$-$&$-$\\
			4 & $5f^{14}7p$&$^2${P}$\rm ^o$ &{1/2} &50770&$-$&$-$&$-$\\
			5 & $5f^{14}7p$&$^2${P}$\rm ^o$ &{3/2} &75719&$-$&$-$&$-$\\
			6 & $5f^{14}8s$&$^2${S}&{1/2} &127703&$-$&$-$&$-$\\
			
		\end{tabular}
	\end{ruledtabular}
\end{table}

Calculated energy levels of Hf and its first three ions are presented in Table~\ref{t:HfEL} and are compared with the experiment.
Good agreement between the sets of data indicates that applied approximation is sufficiently accurate to proceed to the calculations of the hyperfine structure. Energy levels of Hf and Hf$^+$ were calculated before (see, e.g. ~\cite{Dzuba2} for Hf and \cite{Rf+} for Hf$^+$).
We do not make a  direct comparison between the results because to assess the accuracy of the method, it is sufficient to compare the result with  the experiment.
However, it is useful to understand the reasons for some differences in our results with the results of previous calculations of Ref.~\cite{Dzuba2}.
Some energy levels calculated in ~\cite{Dzuba2} are closer to the experiment than in the present work (e.g., low energy states); others (e.g., some high energy states) are closer to the experiment in our present work. The main reason for the  differences is the use of the different versions of the CI+SD method.
The method of Ref.~\cite{CI+SDS} was used in Ref.~\cite{Dzuba2}, while in the present work, we use the method of Ref.~\cite{Dzuba_SD+CI}.
Another reason for the differences comes from the fact that in the present work we do not include radiative corrections. 
This is because  we focus mostly on the hyperfine structure. However, the method of inclusion of the radiative corrections developed in Ref.~\cite{QED} and used in Ref.~\cite{Dzuba2} is applicable for the energy levels and transition amplitudes but not applicable for the singular operators like the operators of hfs.

Calculated energy levels of Rf and its first three ions are presented in Table~\ref{t:RfEL} and compared with other calculations.
Energy levels of neutral Rf were calculated in a number of earlier works~\cite{Dzuba2,Mosyagin,Eliav,Dzuba},  energy levels of Rf+ were calculated in Refs.~\cite{Eliav,Dzuba2}, only the ionization potential (IP) of Rf~III and Rf~IV were reported before~\cite{Johnson,Dzuba2,Dzuba}. The origin of the  differences in the energies of Rf in our present work and earlier work of Ref.~\cite{Dzuba2} is the same as for Hf; see discussion above.

As can be seen from the table, the results of the present calculations for Rf and Rf$^{+}$ are in excellent agreement with previous studies; the difference between the energies of earlier works and present results for Rf is within 300 cm$^{-1}$ for a majority of energy levels, and it is up to $\sim$~1000 cm$^{-1}$ for some states. 
The difference for Rf$^+$ is within $\sim$~2000 cm$^{-1}$, and for some states it is significantly smaller.


 
 The ionization potentials for Rf~III and Rf~IV have been  calculated, and the results obtained are 192367 cm$^{-1}$ and 257396 cm$^{-1}$, respectively. Those results are compared with experiment and other theoretical studies. In Ref.~\cite{Johnson}, the measured results  achieved for the IP are 191960 cm$^{-1}$ and 257290 cm$^{-1}$ for Rf~III and Rf IV, respectively; and in Ref.~\cite{Dzuba}, the calculated results obtained are 192301 cm$^{-1}$ and 257073 cm$^{-1}$, respectively. All these values are in excellent agreement with the results of the present work. For Rf~III, the difference is just 407 cm$^{-1}$ compared with~\cite{Johnson} and 66 cm$^{-1}$ compared with~\cite{Dzuba}; and for Rf~IV, the variation is just 106 cm$^{-1}$ compared with~\cite{Johnson} and 323 cm$^{-1}$ compared with~\cite{Dzuba}. This  is well inside of the error bars of this work.

Comparison of the spectra of Rf and its ions (Table~\ref{t:RfEL}) with the spectra of Hf and its ions (Table~\ref{t:HfEL}) show many similarities and some differences. The most prominent difference is the difference in the ground state configurations of the double and triple ionized ions. This difference comes from the relativistic effects, which pull $s$-electrons closer to the nucleus, reversing the order of the $7s$ and $6d$ states of the Rf ions on the energy scale compared to the $6s$ and $5d$ states of the Hf ions.  

\subsection{Hyperfine structure of Hf~I and Hf~II.}

\label{s:PC}

As it was discussed in sections \ref{s:Ba+}, calculation of the hfs in cases when $d$-states are involved often leads to poor accuracy of the results.
This is because the density of the $d$-states in the vicinity of the nucleus is negligible, and all values of the hfs constants come from higher-order corrections, which include mixing with $s$-states. If leading higher-order corrections are included, then the accuracy for some states might be sufficiently good. It is important to have a way of recognizing such states. Then we would be able to recommend which states of Rf or its ions should be used to extract nuclear moments from the comparison of the measured and calculated hfs. It was suggested in section \ref{s:Ba+} to study partial contributions to the hfs matrix elements. It is also important to study the relative values of the higher-order corrections. In this section, we perform such  study for magnetic dipole and electric quadrupole hfs constants of Hf and Hf$^+$ for the states where experimental data are available. 
Table~\ref{t:DCAHf} shows leading and higher-order contributions to the magnetic dipole hfs constants $A$ for five even and two odd states of Hf.
Table~\ref{t:PWAHf} shows partial wave contributions to the same $A$ constants of the seven states of Hf. 
Leading and higher-order contributions to the magnetic dipole hfs constants $A$ for five even and two odd states of Hf.
Tables ~\ref{t:DCBHf} and ~\ref{t:PWBHf} show similar data for the electric quadrupole hfs constant $B$.

Studying Tables~\ref{t:DCAHf}, \ref{t:PWAHf}, \ref{t:DCBHf}, \ref{t:PWBHf} as well as Table~\ref{t:Ba+} reveal that accurate  calculated values of the hfs constants are likely to be found for the states which satisfy three conditions:
\begin{itemize}
\item The value of the hfs constant is relatively large;
\item There is no strong cancellation between different contributions;
\item The value of the hfs constant is dominated by partial contributions from the low angular momentum states.
\end{itemize}
As one can see, only magnetic dipole hfs of the ground state of Hf fully satisfies these conditions. The difference between theory and experiment, in this case, is about 3\%. In all other cases, including the electric quadrupole hfs constant of the ground state, there is  a large contribution from the 4$d_{5/2}$ channel. However, the accuracy of the results is reasonably good for both types of the hfs constants for some other states as well. This means that the conditions above are rather most favorable than necessary conditions. On the other hand, all cases with poor results can be explained by strong cancellation between different contributions and large contribution from the $d_{5/2}$ partial wave.

Studying Tables~\ref{t:Ba+}, \ref{t:DCAHf}, \ref{t:PWAHf}, \ref{t:DCBHf}, \ref{t:PWBHf} also allows to find a way of  a rough estimation of the uncertainty of the calculations and assign specific error bars to all theoretical results. Dominating contribution to the error usually comes from the contributions of the $d_{3/2}$ and $d_{5/2}$ partial ways. The data in Table~\ref{t:Ba+} shows that the error for the $d_{3/2}$ contribution is about 10\%, while the error for the $d_{5/2}$ contribution is about 20\%. The contribution of the other partial ways to the error budget can be neglected because of either a small error ($s$ and $p$ waves) or  a small contribution. It is natural to assume that the accuracy of the non-diagonal and two-particle contributions is also $\sim$ 10\% since both these contributions have matrix elements with $d$ states. Then the total error for a state $a$ can be calculated as
\begin{equation}\label{e:er}
\sigma_a =\sqrt{\sigma_{an.d.}^2+\sigma_{ad_{3/2}}^2+\sigma_{ad_{5/2}}^2+\sigma_{aTP}^2},
\end{equation}
where $\sigma_{an.d.} = \delta_{an.d.}/10$, $\sigma_{ad_{3/2}} = \delta_{ad_{3/2}}/10$, $\sigma_{ad_{5/2}} = \delta_{ad_{5/2}}/5$, $\sigma_{TP} = \delta_{TP}/10$. Here $\delta$ stands for a particular contribution. The values of $\delta$ can be found in Tables~\ref{t:DCAHf}, \ref{t:PWAHf}, \ref{t:DCBHf}, \ref{t:PWBHf}. Error bars for the hfs constants of Hf, calculated using (\ref{e:er}), are presented in Tables~\ref{t:DCAHf} and \ref{t:DCBHf}.
One can see that in many cases estimated error bars are larger than the actual difference between theory and experiment. However, in some cases of externally strong cancellations between different contributions, the estimated error bars are smaller than the difference between theory and experiment.
This probably means that 
 such states should be excluded from the consideration.

\begin{table}
\caption{Contributions to the magnetic dipole hfs constants of $^{179}$Hf (MHz). The CI values correspond to formula (\ref{e:mex}); HS is the difference between  (\ref{e:mez}) and (\ref{e:mex}); SR is structure radiation (Fig.~\ref{f:SR}); SE is self-energy corrections (Fig.~\ref{f:SE}); TP is two-particle correction (Fig.~\ref{f:TP}). The {\em Sum} line contains the sums of all single-particle contributions (CI,HS,SR,SE). The {\em Total} line contains also TP contributions. The {\em Final} line contains error bars calculated according to (\ref{e:er}).
Experimental values are taken from \cite{Hf-hfs1,Hf-hfs2,Hf-hfs3}.}
\label{t:DCAHf}
\begin{ruledtabular}
\begin{tabular}{l rrrr rrr}
&\multicolumn{1}{c}{$^3$F$_2$}&
\multicolumn{1}{c}{$^3$F$_3$}&
\multicolumn{1}{c}{$^3$F$_4$}&
\multicolumn{1}{c}{$^1$D$_2$}&
\multicolumn{1}{c}{$^3$P$_2$}&
\multicolumn{1}{c}{$^3$D$_2^{\rm o}$}&
\multicolumn{1}{c}{$^5$G$_2^{\rm o}$}\\
\hline
CI    & -82.18 & -35.24 & -16.04 & -32.38 & -41.36 & -42.18 & 191.16 \\
HS    &  15.34 &   5.70 &   2.90 &  14.29 &  15.24 &   12.22 & -52.21 \\
SR    & -13.08 & -16.19 & -19.67 & -18.35 & -13.87 & -16.88 & -17.80 \\
SE    &  -0.90 &   1.70 &   2.66 &   3.13 &   3.28 &   2.50 & -13.30 \\
Sum   & -80.82 & -44.03 & -30.15 & -33.31 & -36.71 & -44.34 & 107.85 \\
TP    &  11.33 &  -8.18 & -12.69 & -19.34 & -25.55 &  -4.92 &  37.17 \\
Total & -69.49 & -52.21 & -42.84 & -52.65 & -62.26 & -49.26 & 145.02 \\
Final & -69(7) & -52(9) & -42(9) & -52(9) & -62(14) & -49(6) & 145(33) \\
\hline
Expt. & -71.43 & -50.81 & -43.46 & -47.68 & -44.7~ & -46.93 & 128.74 \\
\end{tabular}
\end{ruledtabular}
\end{table}


\begin{table}
\caption{Contributions of different partial waves into the magnetic dipole hfs constants of $^{179}$Hf (MHz). The n.d. stands for non-diagonal contributions, which include $s_{1/2} - d_{3/2}$, $p_{1/2} - p_{3/2}$, $d_{3/2} - d_{5/2}$, etc. contributions. The TP terms are not included.}
\label{t:PWAHf}
\begin{ruledtabular}
\begin{tabular}{l rrrr rrr}
&\multicolumn{1}{c}{$^3$F$_2$}&
\multicolumn{1}{c}{$^3$F$_3$}&
\multicolumn{1}{c}{$^3$F$_4$}&
\multicolumn{1}{c}{$^1$D$_2$}&
\multicolumn{1}{c}{$^3$P$_2$}&
\multicolumn{1}{c}{$^3$D$_2^{\rm o}$}&
\multicolumn{1}{c}{$^5$G$_2^{\rm o}$}\\
\hline
 n.d.         & -37.22 &   0.37 &  23.34 &  27.19 &   8.46 &  -6.52 & -64.55 \\
$s_{1/2}$ &  18.89 &  -5.31 & -12.34 &  23.69 &  27.71&  12.46 &-206.51 \\
$p_{1/2}$ &   0.42  &   0.28 &   0.12 &   0.24 &   0.32  & -31.63 &  58.40 \\
$p_{3/2}$ &   0.38  &   0.16 &   0.10 &  -1.09 &  -0.31  &   1.94 &   3.61 \\
$d_{3/2}$ & -75.17 &   5.78 &   5.60 & -46.41 &  -2.95 & -41.13 & 177.65 \\
$d_{5/2}$ &   7.92  & -44.32 & -44.20& -36.97& -69.08 &  20.59 & 137.18 \\
$f_{5/2}$  &   2.58  &   0.25 &  -0.57 &  -0.27 &  -0.24   &   0.09 &   1.45 \\
$f_{7/2}$  &   0.92  &  -1.01 &  -1.75 &  -0.62 &  -0.45   &  0.04 &   0.25 \\
$g_{7/2}$  &  0.36  &    0.03 &  -0.11 &    0.04 &  0.01   & -0.06  &  0.35  \\
$g_{9/2}$  &  0.12  &  -0.27  &   0.36 &  -0.11  &  -0.18  &  0.03  &  0.01  \\
\hline
  Total       & -80.82 & -44.03 & -30.15 & -33.31 & -36.71& -44.34 & 107.85 \\
\end{tabular}
\end{ruledtabular}
\end{table}

\begin{table}
\caption{Contributions to the electric quadrupole hfs constants of $^{179}$Hf (MHz). The meaning of the contribution titles are the same as in Table~\ref{t:DCBHf}.
Experimental values are taken from \cite{Hf-hfs1,Hf-hfs2,Hf-hfs3}. Experimental values are rounded to the last digit before decimal point. More accurate numbers together with error bars can be found in Refs.~\cite{Hf-hfs1,Hf-hfs2,Hf-hfs3}.}
\label{t:DCBHf}
\begin{ruledtabular}
\begin{tabular}{l rrrr rrr}
&\multicolumn{1}{c}{$^3$F$_2$}&
\multicolumn{1}{c}{$^3$F$_3$}&
\multicolumn{1}{c}{$^3$F$_4$}&
\multicolumn{1}{c}{$^1$D$_2$}&
\multicolumn{1}{c}{$^3$P$_2$}&
\multicolumn{1}{c}{$^3$D$_2^{\rm o}$}&
\multicolumn{1}{c}{$^5$G$_2^{\rm o}$}\\
\hline
CI    &  708   &  921   & 1972   & -1107 & -1358 &   153  & 2455 \\
HS    & -108   & -193   & -274   &   153 &   252 &  -155  & -310 \\
SR    &  130   &   49   &   33   &  -115 &   -52 &    99  &  162 \\ 
SE    &    4   &  -12   &  -48   &     9 &    19 &   -11  &    7 \\
Sum   & 734 &  765   & 1683   & -1060 & -1139 &   86   &  2314 \\
TP    & -2  &   17   &   57   & -0.63 &   -22 &   28   &    80 \\
Final & 731 &  783   & 1740   & -1061 & -1162 &   114   &  2394 \\
Error bar & (43) & (144) & (725) & (178) & (243) & (215) & (143) \\
\hline
Expt. & 706 & 931 & 1619 & -905 & -1364 &  740 & 2802 \\
\end{tabular}
\end{ruledtabular}
\end{table}

\begin{table}
\caption{Contributions of different partial waves into the electric quadrupole hfs constants of $^{179}$Hf (MHz). The n.d. stands for non-diagonal contributions, which include $s_{1/2} - d_{3/2}$, $p_{1/2} - p_{3/2}$, $d_{3/2} - d_{5/2}$, etc. contributions. The TP terms are not included.}
\label{t:PWBHf}
\begin{ruledtabular}
\begin{tabular}{l rrrr rrr}
&\multicolumn{1}{c}{$^3$F$_2$}&
\multicolumn{1}{c}{$^3$F$_3$}&
\multicolumn{1}{c}{$^3$F$_4$}&
\multicolumn{1}{c}{$^1$D$_2$}&
\multicolumn{1}{c}{$^3$P$_2$}&
\multicolumn{1}{c}{$^3$D$_2^{\rm o}$}&
\multicolumn{1}{c}{$^5$G$_2^{\rm o}$}\\
\hline
 n.d.     & 865 &-125 &-1819 & -454 &  144 &-1117 & 2525 \\
$p_{3/2}$ &  -5 &   7 &   26 &   31 &   -7 &   44 &  125 \\
$d_{3/2}$ &  70 & 159 & -274 &  294 &  -37 &  108 &  -19 \\
$d_{5/2}$ &-206 & 715 & 3623 & -875 &-1215 &  1041 & -340 \\
$f_{5/2}$ &  39 & 281 &  117 &  418 &   43 &   24 &   54 \\
$f_{7/2}$ & -56 &-273 &  -26 & -464 &  -80 &  -17 &  -51 \\
$g_{7/2}$ &  35 &  48 &   58 &   77 &   36 &    0 &   28 \\
$g_{9/2}$ & -10 & -48 &  -22 &  -87 &  -22 &    3 &   -8 \\
\hline
Sum       & 732 & 765 & 1682 &-1059 &-1139 &   86 & 2314 \\
\end{tabular}
\end{ruledtabular}
\end{table}

\begin{table}
\caption{Contributions into the magnetic dipole hfs constants of the $^2$D$_{3/2}$ and $^4$F$_{5/2}^{\rm o}$ states of the $^{179}$Hf$^+$ ion (MHz). Experimental values are taken from \cite{Exp.Hf+}.}
\label{t:Hf+}
\begin{ruledtabular}
\begin{tabular}{l rr|lrr}
&\multicolumn{1}{c}{$^2$D$_{3/2}$}&
\multicolumn{1}{c}{$^4$F$_{5/2}^{\rm o}$}&
&\multicolumn{1}{c}{$^2$D$_{3/2}$}&
\multicolumn{1}{c}{$^4$F$_{5/2}^{\rm o}$}\\
\hline
   &       &   &   n.d.     &  30.66 &    19.71 \\
CI+HS&  2.11   &  -521.50 &  $s_{1/2}$ & -23.63 &   105.68 \\
SR      &  -28.38   &    -0.02 &  $p_{1/2}$ &  -0.68 &    38.61 \\
SE      &   -1.79   &    23.80 &  $p_{3/2}$ &   1.09 &     1.57 \\
Sum     &  -28.06   &  -497.72 &  $d_{3/2}$ & -64.44 &  -667.02 \\
TP      &   24.06   &   -58.25 &  $d_{5/2}$ &  27.67 &     4.72 \\
Total   &   -4.00   &  -555.97 &  $f_{5/2}$ &   1.06 &    -0.89 \\
Final  & -4(9)&  -556(67) &  $f_{7/2}$ &   0.07 &     0.00 \\
\hline
Expt.   & -17.5(0.9)&  -540(2)      &    Total   & -28.06 &  -497.71 \\

\end{tabular}
\end{ruledtabular}
\end{table}

\begin{table}
\caption{Contributions into the electric quadrupole hfs constants of the $^2$D$_{3/2}$ and $^4$F$_{5/2}^{\rm o}$ states of the $^{179}$Hf$^+$ ion (MHz). Experimental values are taken from \cite{Exp.Hf+}.}
\label{t:Hf+B}
\begin{ruledtabular}
\begin{tabular}{l rr|lrr}
&\multicolumn{1}{c}{$^2$D$_{3/2}$}&
\multicolumn{1}{c}{$^4$F$_{5/2}^{\rm o}$}&
&\multicolumn{1}{c}{$^2$D$_{3/2}$}&
\multicolumn{1}{c}{$^4$F$_{5/2}^{\rm o}$}\\
\hline
           &           &                &        n.d.     &  -342 &  -3228   \\
CI+HS&   1780    &   -897  &  $p_{3/2}$  &       6 &  -36  \\
SR      &      63    &        51 &  $d_{3/2}$ & 2071 &  2301   \\
SE      &     -32      &     -35 &  $d_{5/2}$ &    -9 &   42   \\
Sum     &  1811    &  -881  & $f_{5/2}$ &   34 &  21 \\
TP      &       12     &  -15    &  $f_{7/2}$ &   11 &   -3   \\
Total   &   1823     &   -896  &  $g_{7/2}$ &  40 &   23  \\
Final   & 1823(208)& -896(281) & $g_{9/2}$ &   1 &  -1    \\
\hline
Expt.   & 1928(21)& -728(17)   &    Total    & 1181 &  -881 \\
\end{tabular}
\end{ruledtabular}
\end{table}

We found experimental data on the hfs of Hf$^+$ for only two states, the ground state $5d6s^2 \ ^2$D$_{3/2}$, and the excited odd state $5d6s6p \ ^4$F$^{\rm o}_{5/2}$~\cite{Exp.Hf+}. Calculated contributions to the hfs of these states are presented in Table~\ref{t:Hf+} for the magnetic dipole hfs and Table~\ref{t:Hf+B} for the electric quadrupole hfs. One can see from Table~\ref{t:Hf+} that calculated $A$ hfs constant of the ground state is consistent with zero due to strong cancellation between different contributions. On the other hand, the accuracy of the result for the excited state is high; the difference between theory and experiment is about 3\%. This state satisfies all "most favorable" conditions discussed above. 

For the electric quadrupole hfs constant $B$, the situation is opposite (see Table~\ref{t:Hf+B}), the accuracy is high for the ground state, and it is not so high for the excited state. The latter can be explained by strong cancellation between the non-diagonal contributions and the contributions from the $d_{3/2}$ partial wave.

\subsection{Hyperfine structure of Rf~I and Rf~II}

Hyperfine structure constants $A$ and $B$ calculated for selected states of Rf~I and Rf~II are presented in Table~\ref{t:Rfhfs}.
We have calculated the hfs only for the ground state and for the low-lying states of opposite parity, which are connected to the ground state by electric dipole transitions. The frequencies of these transitions, together with the hyperfine structure, are likely to be measured first.
The  same method of calculations and the same analysis of the partial contributions and error bars were used for Rf~I and Rf~II as for Hf~I and Hf~II (see the previous section). We do not present tables of partial contributions for Rf~I and Rf~II to avoid overloading the paper with technical details. 
Only final results, together with the error bars, are presented in Table~\ref{t:Rfhfs}. As in the case of Hf~I and Hf~II the actual
error of the calculations 
might be significantly smaller than the estimated uncertainties. Note that strong relativistic effects also pull the $7s$ and $7p_{1/2}$ electrons of Rf closer to the nucleus, enhancing their contribution to the hyperfine structure. This might be another reason for a higher accuracy of the calculation for Rf~I and Rf~II compared to what we had for Hf~I and Hf~II.

\begin{table}
\caption{Hyperfine stricture constants $A$ and $B$ for Rf~I and Rf~II. Numeration of states corresponds to Table~\ref{t:RfEL}.
Error bars are calculated with the use of Eq.~(\ref{e:er}).}
\label{t:Rfhfs}
\begin{ruledtabular}
\begin{tabular}{c ll c rrr}
\multicolumn{1}{c}{No.}& 
\multicolumn{1}{c}{Conf.}&
\multicolumn{1}{c}{Term}&
\multicolumn{1}{c}{\textit{J}}&
\multicolumn{1}{c}{Energy}&
\multicolumn{1}{c}{$A/g_I$}&
\multicolumn{1}{c}{$B/Q$}\\
&&&&\multicolumn{1}{c}{cm$^{-1}$}&
\multicolumn{1}{c}{MHz}&
\multicolumn{1}{c}{MHz}\\
\hline
\multicolumn{7}{c}{\textbf{Rf~I}}\\

1  &$7s^{2}6d^{2}$ & $^3${F}     &2 &    0 &  196(108) & 283(18) \\

7  &$7s^{2}7p6d$  & $^3${D}$^{o}$&1 & 8028 &-2546(386) &  716(50) \\
13 &$7s^{2}7p6d$  & $^3${P}$^{o}$&1 &16017 &  788(259) & -738(109) \\
24 &$7s6d^{2}7p$  & $^5${F}$^{o}$&1 &25821 &-14460(1177)& -283(14) \\
                                              
2  &$7s^{2}7p6d$  & $^3${F}$^{o}$&2 & 2737 & 3185(485) &  963(102) \\
8  &$7s^{2}7p6d$  & $^3${D}$^{o}$&2 &11235 & -423(418) &  446(129) \\
10 &$7s^{2}7p6d$  & $^1${D}$^{o}$&2 &13811 &  -40(150) &  578(37) \\
                                              
9  &$7s^{2}7p6d$  & $^3${F}$^{o}$&3 &11328 & 3229(423) & 1015(162) \\
14 &$7s^{2}7p6d$  & $^3${D}$^{o}$&3 &17367 &  939(273) & 1139(217) \\
20 &$7s6d^{2}7p$  & $^5${G}$^{o}$&3 &22941 & 5245(711) &  776(54) \\
\hline

\multicolumn{7}{c}{\textbf{Rf~II}}\\

1  & $7s^{2}6d$& $^2${D}  &{3/2}  &0    &  -190(137) & 1448(130)\\

5  & $7s^{2}7p$& $^2$P$^o$&{1/2}  &19050 & 12960(280) & 0 \\
14 & $7s7p6d$ & $^4$D$^o$ &{1/2} &37378 &-21730(2636) & 0 \\

13 & $7s^{2}7p$& $^2$P$^o$ &{3/2} &33621 & -7798(849) & 2078(113)\\
15 & $7s7p6d$ & $^2$P$^o$ &{3/2} &40015 &  7193(937) & 1414(28)\\
10 & $7s7p6d$ & $^4$F$^o$ &{3/2} &30264 &  9650(984) & 1374(43)\\

12 & $7s7p6d$ & $^4$F$^o$ &{5/2} &33320 & 15810(1828) & 417(141)\\
16 & $7s7p6d$ & $^4$D$^o$ &{5/2} &40640 &  7582(1532) & 1421(259)\\
\end{tabular}
\end{ruledtabular}
\end{table}

\section{Conclusion}
	
In this paper, the energy levels and the hyperfine structure constants $A$ and $B$ for low-lying states of the Rf atom and ions were calculated. 
Energy levels were calculated for Rf~I, Rf~II, Rf~III, and Rf~IV, while hyperfine  structures constnats were calculated for Rf and Rf~II.
Similar calculations were performed for  the  lighter analog of Rf, the Hf atom, and its ions to control the accuracy of the calculations.
Present results are in good agreement with other calculations and previous measurements where the data are available. 
The way of estimation of the uncertainty of the hfs calculations is suggested. For majority  of the states, the uncertainty is within 10\%.
The calculated hfs constants of Rf~I and Rf~II can be used  to extract  nuclear magnetic and electric quadrupole moments from future measurements.
	
\section{Acknowledgements}
This work was supported by the Australian Research
Council Grants No. DP190100974 and DP200100150. This research includes computations using the computational cluster Katana supported by Research Technology Services at UNSW Sydney~\cite{Katana}.

\end{document}